\documentclass[conference]{IEEEtran}
\IEEEoverridecommandlockouts
\usepackage{cite}
\usepackage{comment}
\usepackage{amsmath,amssymb,amsfonts}
\usepackage{algorithmic}
\usepackage{graphicx}
\usepackage{textcomp}
\usepackage{xcolor}
\usepackage{todonotes}
\usepackage{hyperref}
\usepackage{float}

\def\BibTeX{{\rm B\kern-.05em{\sc i\kern-.025em b}\kern-.08em
    T\kern-.1667em\lower.7ex\hbox{E}\kern-.125emX}}
    
\begin{document}

\title{CharmFL: A Fault Localization Tool for Python}

\author
{

\IEEEauthorblockN{Qusay Idrees Sarhan\textsuperscript{1,2},
Attila Szatm\'ari\textsuperscript{1},
Rajmond T\'oth\textsuperscript{1},
\'Arp\'ad Besz\'edes\textsuperscript{1}}
\IEEEauthorblockA{\textsuperscript{1} Department of Software Engineering, University of Szeged, Szeged, Hungary}
\IEEEauthorblockA{\textsuperscript{2} Department of Computer Science, University of Duhok, Duhok, Iraq\\
\{sarhan, szatma, beszedes\}@inf.u-szeged.hu} raymonddrakon@gmail.com

}

\maketitle





\begin{abstract}
Fault localization is one of the most time-consuming and error-prone parts of software debugging. There are several tools for helping developers in the fault localization process, however, they mostly target programs written in Java and C/C++ programming languages. While these tools are splendid on their own, we must not look over the fact that Python is a popular programming language, and still there are a lack of easy-to-use and handy fault localization tools for Python developers. In this paper, we present a tool called ``CharmFL'' for software fault localization as a plug-in for PyCharm IDE. The tool employs Spectrum-based fault localization (SBFL) to help Python developers automatically analyze their programs and generate useful data at run-time to be used, then to produce a ranked list of potentially faulty program elements (i.e., statements, functions, and classes). Thus, our proposed tool supports different code coverage types with the possibility to investigate these types in a hierarchical approach. 
The applicability of our tool has been presented by using a set of experimental use cases. The results show that our tool could help developers to efficiently find the locations of different types of faults in their programs.\\
\end{abstract}

\begin{IEEEkeywords}
Debugging, fault localization, spectrum-based fault localization, Python, CharmFL.
\end{IEEEkeywords}

\section{Introduction}

Software systems and applications cover many aspects of our day-to-day activities. However, they are still far from being free of faults. Software faults may cause critical undesired situations including life loss. Therefore, various software fault localization techniques have been proposed over the last few decades including Spectrum-based fault localization (SBFL)~\cite{intro1}. In SBFL, the probability of each program element (e.g., statements) of being faulty is calculated based on program spectra obtained by executing a number of test cases. However, SBFL is not yet widely used in the industry because it poses a number of issues~\cite{intro2}. One of such issues is that most of the SBFL tools currently target programs written in C/C++ and Java. Thus, there is lack in SBFL tools that help developers debug their programs that are written in other programming languages including Python which is considered also as one of the most popular programming languages.

In this paper, we present a tool called 
``CharmFL'' as a plug-in for the PyCharm IDE, a popular Python development platform, to automate the software fault localization process. Our tool utilizes SBFL to assist Python developers in automatically analyzing their programs and producing useful data at run-time that can then be used to generate a ranked list of potentially faulty program elements. To determine whether a statement is faulty or not, developers examine each statement in turn, beginning at the top of the list (the most suspicious element). 
Several experiments with Python projects were conducted to assess the applicability of our tool. 
The results indicate that the tool is useful for locating faults in various types of programs and that it is simple to use. 

The remainder of the paper is organized as follows. Section~\ref{SBFL_Background} briefly introduces the background of SBFL and its main concepts. Section~\ref{Related_Works} presents an overview of the most related works. Section~\ref{Methodology} provides a theoretical overview on the used techniques in our tool. Section~\ref{CharmFL} presents our proposed software fault localization tool. Section~\ref{Use_Cases} discusses the applicability of our tool in different practical contexts. Finally, we provide our conclusions and possible future works in Section~\ref{Conclusions}.

\section{Background of SBFL}
\label{SBFL_Background}

Fault localization is a time consuming part of the software debugging process, therefore the need for automating it is incredibly important.
There are several approaches to implement the process\cite{survey2}, however we focus on SBFL due to its simple but powerful nature, i.e. using only code coverage and test results.
There have been several surveys
written~\cite{survey1, survey2, survey3} and various empirical studies~\cite{empir1, empir2} performed on this topic.

\begin{figure}[h!]
\includegraphics[width=0.5\columnwidth, height=4cm]{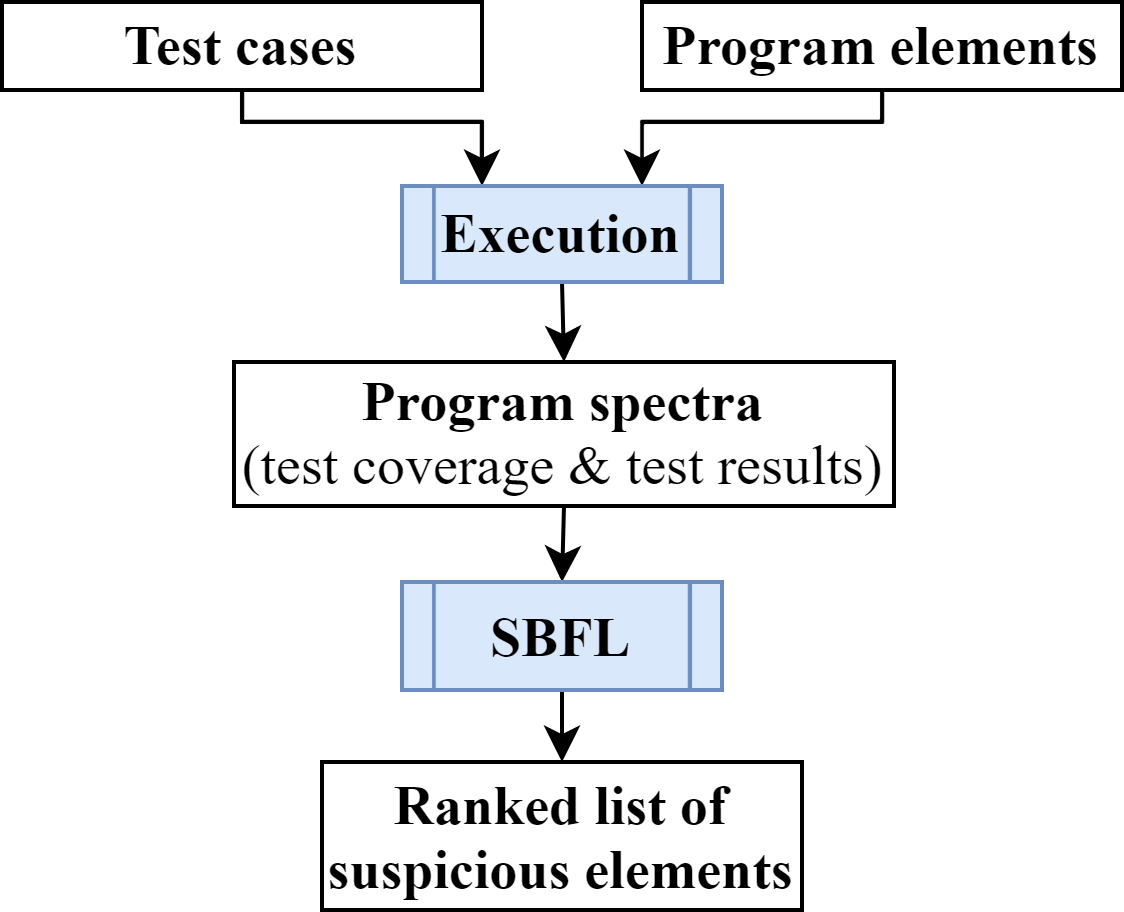}
\centering
\caption{SBFL process}
\label{sbfl_proc}
\end{figure}

Figure~\ref{sbfl_proc} shows the SBFL process. 
Using the program's spectra (i.e., program elements, per-test coverage, and test results), SBFL can help the programmer to find the faulty element in the target program's code easier.
The code coverage matrix is a two-dimensional
matrix used to represent the relationship between the test
cases and the program elements, whose rows demonstrate the test cases and columns represent the program elements. 
An element of the matrix is 1, if it is covered by test case, otherwise it is 0. 
In another matrix vector, the test results are stored where 0 means the test case is passed and 1 when it is failed.
Using these matrices, the following four basic statistical numbers are calculated for each program element $\phi$:
\begin{itemize}
    \item  $\phi_{ep}$: number of passed tests covering $\phi$
    \item  $\phi_{ef}$: number of failed tests covering $\phi$
    \item  $\phi_{np}$: number of passed tests not covering $\phi$
    \item  $\phi_{nf}$: number of failed tests not covering $\phi$\\

\end{itemize}

Then, our tool uses these four numbers with the formula in Equation~\ref{taran}, Tarantula\cite{taran}; Equation~\ref{ochiai}, Ochiai~\cite{metric3}; Equation~\ref{dstar}, DStar\cite{metric2}; Equation~\ref{wong2}, Wong2~\cite{metric4} to provide a ranked list of program elements as an output. 
Whichever element ranked the highest in the list, it is the most suspicious of containing a bug.



\begin{equation}
	\label{taran}
	\begin{split}
		\text{Tarantula}  =  \frac{\frac{\phi_{ef}}{\phi_{ef}+\phi_{nf}}}{\frac{\phi_{ef}}{\phi_{ef}+\phi_{nf}}+\frac{\phi_{ep}}{\phi_{ep}+\phi_{np}}}
	\end{split}
\end{equation}

\begin{equation}
	\label{ochiai}
	\begin{split}
		\text{Ochiai}  =  \frac{\phi_{ef}}{\sqrt{(\phi_{ef}+\phi_{nf}) * (\phi_{ef}+\phi_{ep})}}
	\end{split}
\end{equation}

\begin{equation}
	\label{dstar}
	\begin{split}
		\text{Dstar}  =  \frac{\frac{\phi_{ef}}{\phi_{ef}+\phi_{nf}}}{\frac{\phi_{ef}}{\phi_{ef}+\phi_{nf}}+\frac{\phi_{ep}}{\phi_{ep}+\phi_{np}}}
	\end{split}
\end{equation}

\begin{equation}
	\label{wong2}
	\begin{split}
		\text{Wong2}  =  \phi_{ef} - \phi_{ep}
	\end{split}
\end{equation}

\section{Related Works}
\label{Related_Works}

There are many software fault localization tools implemented and proposed in the literature. This  section  briefly  presents them. Jones et al.~\cite{taran} proposed a standalone software fault localization tool called 
``Tarantula'' to help C programmers to debug their programs. The tool assigns different colors to program statements based on how suspicious they are, ranging from red (most suspicious) to green (not suspicious). Besides, the tool displays varying brightness levels based on how frequently the tests execute a statement. The brightest statements are those that are most commonly executed. However, the tool does not run test cases and record their results; it takes as input a program's source code and the results of executing a test suite on the program. Furthermore, the tool's only supported metric is the Tarantula metric.

Chesley et al.~\cite{rw2} proposed an Eclipse plug-in tool called 
``Crisp'' that helps developers identify the reasons for a failure that occurs due to code edits by constructing intermediate versions of a program that is being edited. For example, if a test case fails, the tool will identify parts of the program that have been changed and caused the failing test. Thus, developers can concentrate only on those affecting changes that were applied.

Ko and Myers~\cite{rw3} proposed a standalone debugging tool called ``Whyline'' for Java programs. The tool employs both static and dynamic slicing to formulate why and why not questions, which are then presented in a graphical and interactive way to help developers in understanding the behavior of a program under test. It also records program execution traces and the status of each used class whether it is executed or not. Using the tool also allows the user to load the execution trace of a program and select a program entity at a specific point during its execution. Then he or she can click on the selected entity to bring up a pop-up window containing a set of questions that include data values gathered during the execution as well as information about the properties of the selected entity.

Hao et al.~\cite{rw4} proposed an Eclipse plug-in tool called ``VIDA'' for programs written in Java. The tool extracts statements hit spectrum from the target programs, executes JUnit tests and based on their results, it calculates suspiciousness. It also provides a list of the ten most suspicious statements as potential breakpoints. It displays the history of breakpoints including the developers’ previous estimates of the correctness of the breakpoint candidates as well as their current suspiciousness. Moreover, it employs colors to distinguish between the developers' estimations, ranging from red (wrong) to green (correct), and suspiciousness, ranging from black (very suspicious) to light gray (less suspicious). And, it provides the users with the ability to extract static dependency graphs from their programs to assist developers with their estimations and also to help them understand the relationships among different program entities.

Janssen et al.~\cite{rw5} and Campos et al.~\cite{rw6} proposed a fault localization tool that adopts SBFL and it is available as a command-line tool called ``Zoltar'' and as an Eclipse plug-in called ``Gzoltar''.  The tool provides a complete infrastructure to automatically instrument the source code of the programs under test in order to generate runtime data, which is then used to return a ranked list of faulty locations. It also uses colors to mark the execution of program entities from red to green based on their suspiciousness scores. The tool only employs the Ochiai metric to compute suspiciousness.

Wang et al.~\cite{rw7} proposed a fault localization tool called ``FLAVS'' for developers using Microsoft Visual Studio platform. The tool provides an automatic instrumentation mechanism to record program spectrum information during the execution. It also provides a user with two options either automatically or manually to mark the result of each used test case; whether it is successful or not. Additionally, it monitors each test environmental factors of the running program such memory consumption, CPU usage, and thread numbers. For example, the developer can notice that there is something wrong when the CPU time drops to zero and never gets increased again during the running of a test case. The tool provides different levels of granularities for fault localization analysis such as statement, predicate, and function. Using the tool allows the users to examine the correct positions in the source code files by clicking on the suspicious units, which are displayed and highlighted in different colors also. The functionalities of ``FLAVS'' have been extended by Chen and Wang~\cite{rw8} in another tool called ``UnitFL''. The tool uses program slicing to decrease the program execution time. Besides, it provides different levels of granularities for fault localization analysis to provide different aspects of execution during the program analysis. And, it shows fault-related elements with different colors based on their suspiciousness; ranging from green to red.

Ribeiro et al.~\cite{rw9} proposed a SBFL tool called ``Jaguar'' for Java developers. The tool supports two advanced spectra types which are control-flow and data-flow. Also, it visualizes suspicious program elements where the user can easily inspect suspicious methods, statements, or variables. Although the data-flow spectrum provides more information, it is not adopted widely in SBFL because of the high costs of execution. To overcome this issue, the tool utilizes a lightweight data-flow spectrum coverage tool called ``ba-dua''. This enables the tool to be used for testing large-scale programs at affordable execution costs. The tool can be used as an Eclipse plug-in or as a command-line tool.



All the previous tools target programs written in Java and C/C++ programming languages. Tools for helping Python developers in their debugging process have not been previously proposed in the literature by other researchers. However, two open-source fault localization tools for Python's pytest testing framework are available, namely, Fault-Localization~\cite{py1} and PinPoint~\cite{py2}. 
In this paper, we propose a tool called ``CharmFL'' with more features to target programs written in Python; which is considered one of the most popular programming languages nowadays. Compared to the other two tools, our proposed tool supports different types of code coverage (i.e., class, method, and statement), displays the fault localization results in different ways, provides a graphical user-friendly interface to examine the suspicious elements, and enabling the user to smoothly examine any suspicious element via clickable links to the source code. Table~\ref{Python_Tools} summarizes the features of our proposed tool compared to the others.



\begin{table}[h!]
\caption{Comparison among Python fault localization tools}
\label{Python_Tools}
 \centering
 \resizebox{1\columnwidth}{!}
 {
 \setlength{\tabcolsep}{3pt}
\begin{tabular}{|l|c|c|p{100pt}|}
\hline
Features &  Fault-Localization&PinPoint&CharmFL \\ 
\hline
Statement hit coverage&  Yes & Yes&Yes \\ 
\hline
Method hit coverage& No& No& Yes \\
\hline
Class hit coverage& No& No& Yes \\
\hline
Supported SBFL metrics& 1  &5 & 4 \\
\hline
Shows ranking&   Color-based& Value-based list&  Both  \\
\hline
Shows suspicious  scores&  Yes & No&Yes \\ 
\hline
Ties ranking& No& No& Min, Max, or Average \\
\hline
GUI interface&  No & No& Yes \\
\hline
Command-line interface&  Yes & Yes& Yes \\
\hline
Elements investigation&   Flat& Flat& Hierarchy  \\
\hline
Elements navigation &   No& No& Via clickable links to each element in the source code  \\
\hline
Tool type&  Option for pytest framework & Option for pytest framework&  Plug-in for PyCharm IDE \\
\hline
Current version&   0.1.6& 0.3.0& 0.1  \\
\hline
 
\end{tabular}
}
\end{table}


\section{Methodology}
\label{Methodology}

\begin{figure}[h!]
\includegraphics[width=0.6\columnwidth, height=6cm]{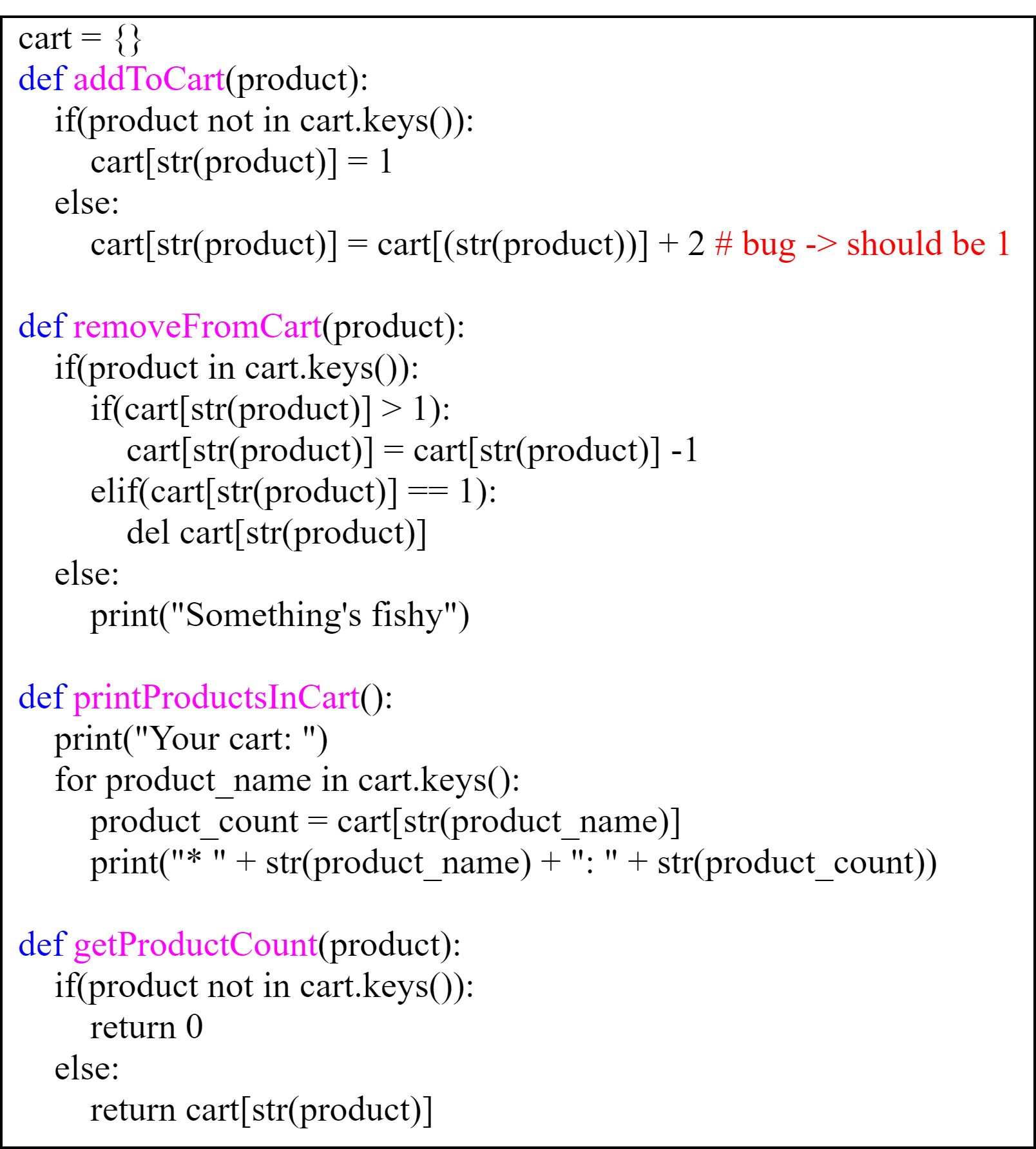}
\centering
\caption{Running example – program code}
\label{code}
\end{figure}

\begin{figure}[h!]
\includegraphics[width=0.6\columnwidth, height=6cm]{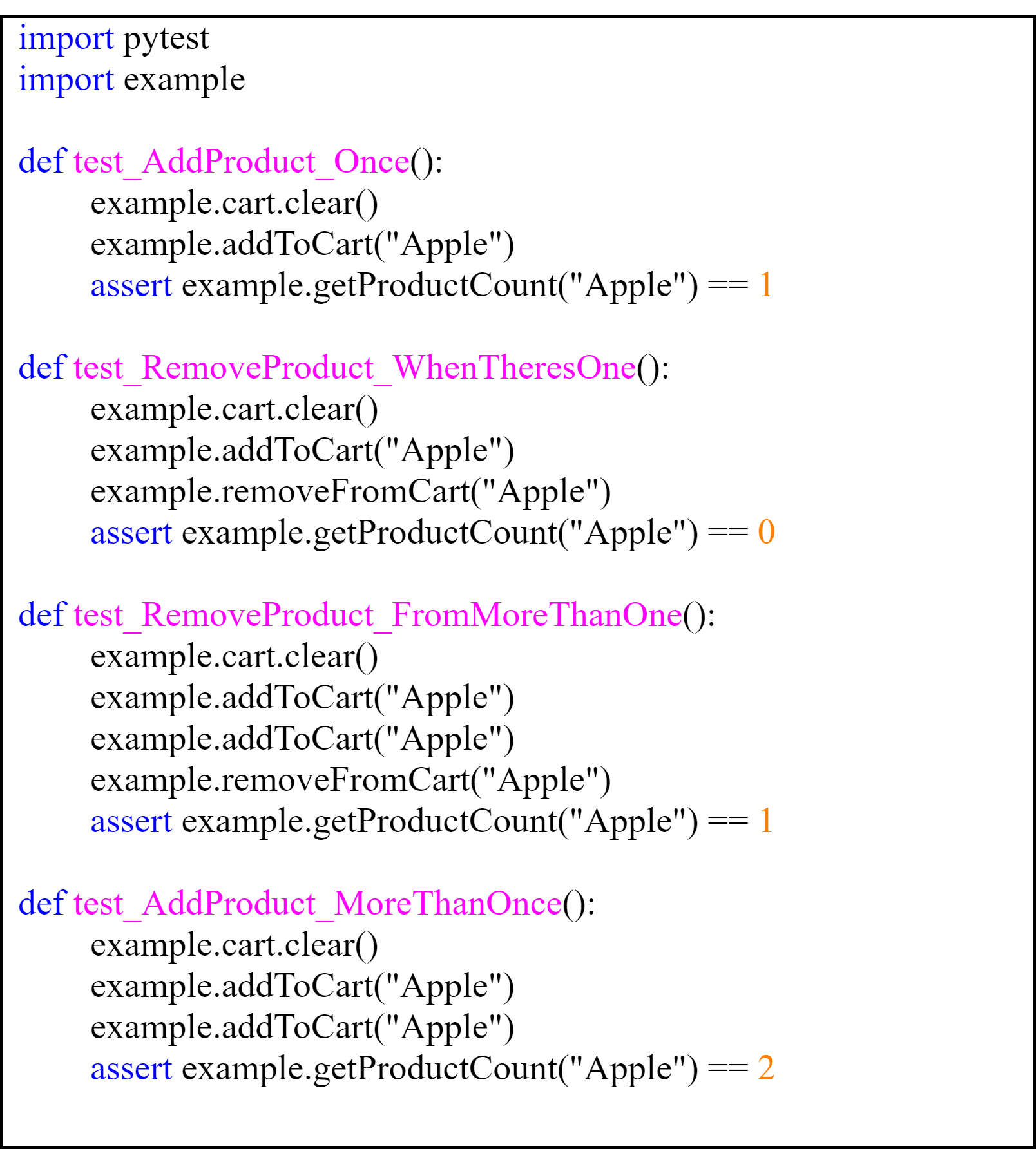}
\centering
\caption{Running example – test cases}
\label{tests}
\end{figure}

In this section, we give a theoretical overview on the used techniques in our tool. 
We will use an example project for demonstration purposes. The selected project has four methods as shown in Figure~\ref{code} and four test cases to test them as shown in Figure~\ref{tests}. For simplicity, we will represent the four test cases through the paper as T1, T2, T3, and T4 according to their order in the figure.

Our tool provides the opportunity to measure statement, method, and class coverage levels. This is achieved by employing the ``zooming in/out''  hierarchy approach, where the user can examine the suspicious elements from the highest level in the hierarchy (i.e., classes) to lower levels in the hierarchy and repeat the steps above, until s/he reaches the lowest level, which is the statements level.
This is better than only one level of granularity as the developer can exclude methods or even classes from the ranking list, thus saving time spent on the debugging process.

We can see in this example that the highest granularity is method level; in this case, the class level coverage is absent.
Table~\ref{cov_matrix} presents the method level coverage matrix, and the basic statistical numbers. 
Running any SBFL algorithm, e.g. Tarantula, we get a list of elements with suspiciousness scores as presented in Table~\ref{scores}. 
For the lack of space, we will not show the statement granularity, but the overarching principle is the same; we investigate the elements with highest scores until we find the bug.
We can see that the ``addToCart'' method has the highest score according to Tarantula. 
Using the ``zooming in/out'' technique, we need to investigate the statements in the ``addToCart'' method first. 
This saves the developers time since they do not have to go through all the statements in the suspicousness list. 

\begin{table}[h]
\caption{Method hit spectrum (with four basic statistics)}
\centering
\label{cov_matrix}
\resizebox{0.8\columnwidth}{!}{%
\begin{tabular}{|l|c|c|c|c||c|c|c|c|}
\hline
 & \multicolumn{1}{l|}{T1} & \multicolumn{1}{l|}{T2} & \multicolumn{1}{l|}{T3} & \multicolumn{1}{l||}{T4} & \multicolumn{1}{l|}{ef} & \multicolumn{1}{l|}{ep} & \multicolumn{1}{l|}{nf} & \multicolumn{1}{l|}{np} \\ \hline
addToCart & 1 & 1 & 1 & 1 & 2 & 2 & 0 & 0 \\ \hline
removeFromCart & 0 & 1 & 1 & 0 & 1 & 1 & 1 & 1 \\ \hline
printProductsInCart & 0 & 0 & 0 & 0 & 0 & 0 & 2 & 2 \\ \hline
getProductCount & 1 & 1 & 1 & 1 & 2 & 2 & 0 & 0 \\ \hline
\hline
Test results & 0 & 0 & 1 & 1 &  &  &  &  \\ \hline
\end{tabular}%
}
\end{table}

\begin{table}[h!]
\caption{Tarantula suspicousness scores}
\centering
\label{scores}
\resizebox{0.4\columnwidth}{!}{%
\begin{tabular}{|l|c|}
\hline
 
Method& Score\\\hline
addToCart & 0.58 \\ \hline
removeFromCart & 0.41 \\ \hline
printProductsInCart & 0.00 \\ \hline
getProductCount & 0.48 \\ \hline
\end{tabular}%
}
\end{table}

\section{CharmFL Tool}

\label{CharmFL}

In this section, we give an overview about our tool's architecture, data processing, and user interface.
Our tool can be divided into two parts; front-end and the back-end framework. The first part is the actual plug-in for the PyCharm IDE, which the user can interact with and use during debugging.
We detail this part in Section~\ref{PyCharm_plugin}. The second part is a framework that gives the opportunity for developers to integrate fault localization in other IDEs. We give details on its architecture and usage in Section~\ref{architecture}.

\subsection{GUI}
\label{PyCharm_plugin}
\begin{figure}[h!]
\includegraphics[width=\columnwidth]{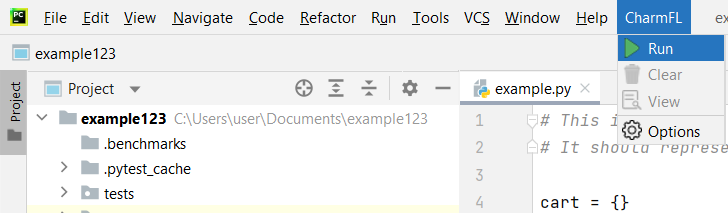}
\caption{CharmFL GUI}
\label{charmfl_gui}
\end{figure}

The front-end part of the tool, shown in Figure~\ref{charmfl_gui}, is an IDE specific plug-in using the CharmFL engine for the PyCharm IDE.
After installing the plug-in and opening the Python project in the IDE, the user can run the fault localization process to get the list of program suspicious elements.

\begin{figure}[h!]
\includegraphics[width=0.9\columnwidth]{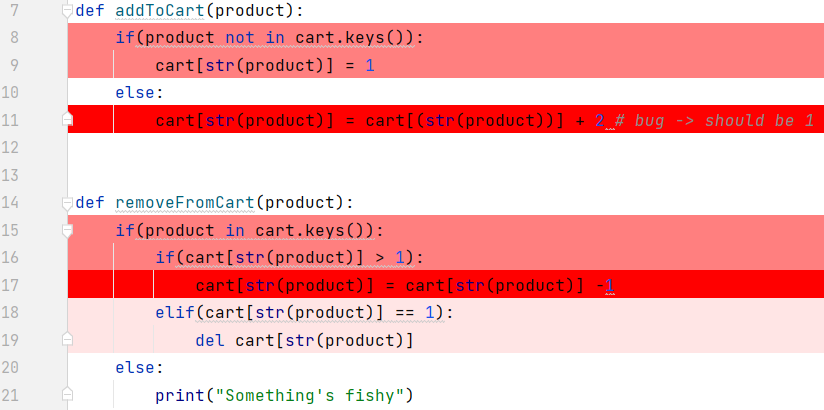}
\caption{ Highlighted statements based on suspicious scores}
\label{charmfl_colors}
\end{figure}

Additionally, the corresponding program elements are highlighted with different shades of red color based on the suspicious scores as shown in Figure~\ref{charmfl_colors}. The darker the color is, the most suspicious the element is.
If the user accidentally closes the results table, s/he can reopen it again by clicking on the View button in the CharmFL menu.

\begin{figure}[h!]
\includegraphics[width=0.3\columnwidth, height=4cm]{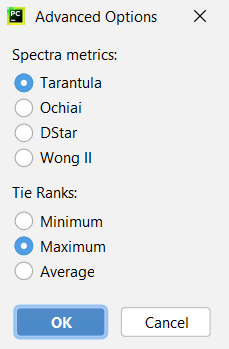}
\centering
\caption{CharmFL advanced options}
\label{advanced}
\end{figure}

There is a set of advanced options for researchers too which appears via clicking the Options button of the menu as shown in Figure~\ref{advanced}. Such options enable them to select different metrics for comparison and to apply different tie-breaking techniques to the elements sharing the same score in the ranking list.

When the user selects multiple metrics, there will be a table for each metric, that way they can compare the elements side-by-side.
This is especially good for researchers who would like to compare the efficiency of the supported SBFL metrics.

\begin{figure}[h!]

\includegraphics[width=0.8\columnwidth]{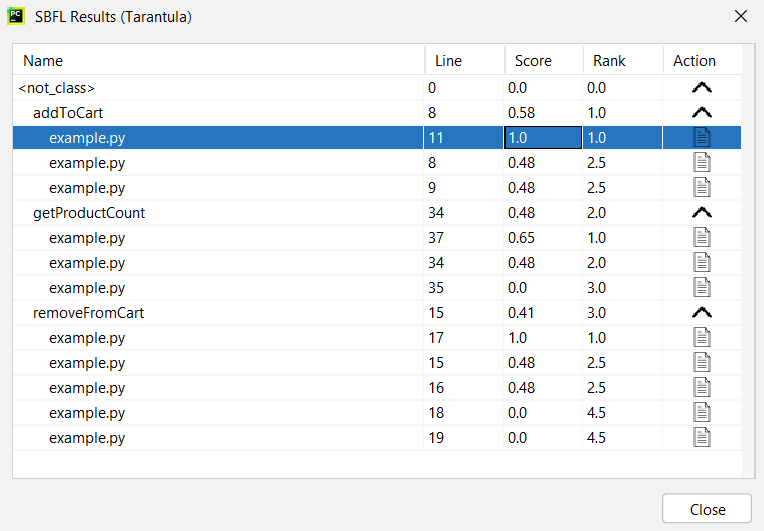}
\centering
\caption{CharmFL rankling list output}
\label{output}
\end{figure}

The SBFL results table (Figure~\ref{output}) shows the program elements hierarchically, next to them there are their positions in the source code, their ranks, and their scores. Also, the Action button can be used to hide/show the elements inside each level of the hierarchy or to jump on a specific element via clicking on its corresponding document icon.

\label{fw}

\subsection{Framework's Architecture}
\label{architecture}

This is the part of the tool where we gather and process the coverage and test result data. 
The framework can be used as a stand-alone tool or integrated in other IDEs too as a plug-in. 


In order to collect the program's spectra, code coverage measurement is needed. To obtain the code coverage, the target program needs to be instrumented.
For this purpose, our tool uses the popular coverage measuring tool for Python, called ``coverage.py''~\cite{py3}.
This tool can measure on either statement or branch  coverage levels, however in its current format it is not able to measure method or class coverage levels. 
Our framework transforms the statement level to method and class levels as shown in Figure~\ref{output}. This is achieved by putting all the statements of each function under the corresponding function's name and then putting all the functions of each class under the corresponding class's name. Thus, each function will has its own set of statements and each class its own set of functions including the statements. Afterward, the classes are sorted based on their suspiciousness scores, then the functions, and finally the statements. For example, the statement at line 37 will not be examined before the statement at line 8 because the latter is belong to a function of higher rank in the ranking list. This hierarchical coverage feature gives additional useful information about the suspicious scores on all layers to the user. 
They can exclude whole methods or even classes based from the list.

\begin{figure}[h!]
\includegraphics[width=0.8\columnwidth,height=1.25cm ]{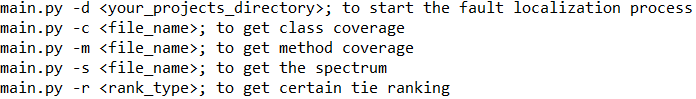}
\centering
\caption{Framework usage}
\label{options}
\end{figure}

Additionally, in order to make the coverage matrix, we used the ``.coveragerc'' file where the user can configure the measurement. 
After collecting the coverage report, we run tests using ``pytest''~\cite{py4} to fetch the results.
Having those collected, we make coverage and test results matrices according to Jones et al.~\cite{taran} from the raw data.
Afterward, the tool calculates the suspiciousness score for each program element based on the equations as described in Section~\ref{SBFL_Background}.
The framework provides class, method, and statement coverage levels; test results; coverage matrix and the hierarchical ``ranking list''.
Figure~\ref{options} lists the usages of the framework.

\section{Potential Use Cases}
\label{Use_Cases}

When programmers face bugs in their Python programs, they have a couple of directions to go with in the debugging life-cycle. 
In order to find the bug, they can either run the test suite to figure out which test is failing and go from there, or they can inject break points in the code to investigate each value while pausing the program execution.

In this section, we will show three ways of how to use our tool in various phases of the debugging, as follows:
\begin{enumerate}
    \item Running the test suite, then start the CharmFL tool.
    \item Running the test suite, inject break points and then start the CharmFL tool.
    \item Running the CharmFL tool, then inject break points.\\
\end{enumerate}

For each scenario, we start from the point when the existence of the bug was first detected, i.e. someone reported the bug while using the software. For demonstration purposes, we will use a simple example project; but also any other Python project can be used. 
The example project has four methods, and four test cases that cover 90\% of the program. 
We injected an artificial bug, in the 11th line, so two of the four test cases would fail.
Our tool is successful if the 11th statement is in the top-10 of the list of suspicious elements, and the debugging is successful if the test cases pass after the bug is fixed.


First, we demonstrate the usage of our tool after running the test suite.
Having done that, we get from pytest's report that there are two failing test cases.
We open the test files, meanwhile we start the CharmFL tool.
Reading the pytest's test results report, we can see that there are two failing test cases. 
Additionally, we can see that in the test cases the method ``addToCart'' is called two times and the ``removeFromCart'' is called once. Hence, we start with the examination of the ``addToCart'' method.
When we investigate the method we can see that there are three statements.
At this point, we start the CharmFL tool and use it to decide which statement to investigate first.
We click on the first element in the method with the highest score in the list of suspicious elements and try to fix the bug in the statement. 
We run the test cases again and see that all test cases pass.
In this scenario, our tool helped deciding which statement should be investigated first, hence saving time on debugging.

Second, we use our tool with a bit more advanced debugging technique; break-point oriented debugging.
The PyCharm IDE has a built-in debugger, which is the best option to use alongside our tool.
First, we run the tests and investigate the failing ones similarly like in the previous scenario. 
Again, we have two failing test cases that cover the ``addToCart'' and ``removeFromCart'' methods.
Then, we insert break points to those lines that are covered by the failing test cases.
Next, start the debugging session to investigate what values do the variables take and what is not going according to the plan.
Meanwhile, we start the CharmFL tool to get the most suspicious elements.
We can see in Figure~\ref{charmfl_colors} that the 11th statement in the example.py file is dark red, which is very suspicious.
We fix the bug in the statement, then verify the fix by running the tests again.
We can conclude from this scenario, that checking against the suspiciousness list can help the programmer a lot with the debugging.


The final scenario, we start the CharmFL tool before doing any debugging. 
In this scenario, we start the tool then look at the list of suspicious elements.

Developers tend to investigate only the first ten (also referred to as top-10) elements in the ranking list because after that, they start to lose interest in using the tool~\cite{TopN1, TopN2}.
Therefore, a fault localization algorithm is successful if it can fit as many faulty elements in the top-10 list as possible. 

Using this technique, we look at and click on the element with the highest score in the suspiciousness list table shown in Figure~\ref{output}.
The tool then redirects us to the statement we want to investigate. 
The background color refers to the suspiciousness level, i.e. how likely the statement is to contain a fault, the darker the color the higher the suspicion is.
We can see in Figure~\ref{charmfl_colors} that the statement has a dark red background, meaning it is the most likely to contain a bug.
When investigating the element we can use break-point-based debugging.
Without running the test cases, we can place a break-point to the statement we clicked on. 
We fix the statement and run the tests for verification.
This scenario takes a few more extra steps. 
However, this is a helpful guide when the test cases are well defined and maintained.
In this case, our tool can reduce the excessive time and energy that would have been spent on debugging.

\section{Conclusions}
\label{Conclusions}


This paper describes ``CharmFL''\footnote{ \url{https://sed-szeged.github.io/SpectrumBasedFaultLocalization/}}, an Open-source fault localization tool for Python programs. 
The tool is developed with many interesting features that can help developers debugging their programs by providing a hierarchical list of ranked program elements based on their suspiciousness scores. The applicability of our tool has been evaluated via different use cases.
The tool has been found to be useful for locating faults in different types of programs and it is easy to use. 
For the future work, we would like to implement interactiveness to enable the user to give his/her feedback on the suspicious elements to help re-rank them, thus improving the fault localization process.
Also, we would like to add other features such as displaying the tool's output using different visualization techniques. Finally, assessing the tool with real users and in real-world scenarios would be a valuable next step too.


\section*{Acknowledgements}
The research was supported by the Ministry of Innovation and Technology, NRDI Office, Hungary within the framework of the Artificial Intelligence National Laboratory Program, and by grant NKFIH-1279-2/2020 of the Ministry for Innovation and Technology. Qusay Idrees Sarhan was supported by the Stipendium Hungaricum scholarship programme.

\bibliographystyle{IEEEtran}
\bibliography{references.bib} 

\vspace{12pt}

\end{document}